\documentclass[showpacs,aps,prl,floats,twocolumn]{revtex4}
\usepackage{bm}
\usepackage{braket}
\usepackage{graphicx}
\usepackage{amsfonts}
\usepackage{amsmath}
\usepackage{amssymb}
\usepackage{color}
\usepackage{ulem}
\usepackage{amsmath,amssymb}
\usepackage[caption=false]{subfig}

\begin{document}

\abovedisplayskip=7pt
\abovedisplayshortskip=0pt
\belowdisplayskip=7pt
\belowdisplayshortskip=7pt

\title{Graphene as a tunable THz reservoir for shaping the Mollow triplet of an artificial atom via plasmonic effects}

\author{Ebrahim Forati}
\email {eforati@uwm.edu}
\author{George W. Hanson}
\email {george@uwm.edu}
\address{Department of Electrical Engineering, University of Wisconsin-Milwaukee, Milwaukee, WI 53211, USA}

\author{Stephen Hughes}
\email {shughes@physics.queensu.ca}
\address{Department of Physics, Engineering Physics and Astronomy, Queen's University, Kingston, Ontario, Canada K7L3N6}

\begin{abstract}
Using a realistic quantum master equation we show that the resonance fluorescence spectra 
of a two-level artificial atom (quantum dot) can be tuned by adjusting its photonic local density of states via biasing of one 
or more graphene monolayers. The structured photon reservoir is  included using a photon Green 
function theory which fully accounts for the loss and dispersion. The 
field-driven Mollow triplet spectrum can be actively controlled by the graphene bias in the THz frequency regime. We 
also consider the effect of a dielectric support environment, and multiple graphene layers, on the emitted fluorescence.  
Finally, thermal bath effects are considered and shown to be  important for low THz frequencies.
\end{abstract}

\pacs{42.50.Pq, 78.67.Bf, 73.20.Mf}

\maketitle




\section{Introduction}
It is highly desirable to  electronically manipulate the photonic spectrum of a multi-level emitter such as an atom or quantum dot (QD). While it is well-known that the spectrum is influenced by the photon emitter's electromagnetic environment (e.g., via the Purcell effect \cite{Purcell}), engineering the environment to obtain desirable characteristics often results in a fixed structure that is not actively tunable. Surface plasmon polaritons (SPPs) on graphene \cite{Wunsch}-\cite{Han2008} are highly tunable, and offer a promising way to achieve electronic control over an emitter's spectrum  through interactions with graphene SPPs. Recent scattering-type scanning near field optical microscopy (SNOM) imaging experiments \cite{Chen} have demonstrated in real space the excitation of graphene SPPs on finite graphene structures, and in \cite{Yan} excitation and damping of SPPs on graphene structures was investigated experimentally for graphene on several substrates. Additionally, graphene quantum plasmonics has been considered in \cite{Koppens} where vacuum Rabi splitting was shown, and in \cite{Man} where active control over a quantum state via biasing was demonstrated.

When placed in the vicinity of a multi-level emitter, graphene, along with the vacuum density of electromagnetic field modes, forms the photonic reservoir with which the emitter interacts. The spectral and statistical properties of such a system is strongly dependent on the reservoir mode density via the local density of states (LDOS) \cite{Carmichael1}. In \cite{Vlack}, the reservoir of electromagnetic modes is altered by the presence of a metal nano-particle, and the resonance fluorescence was examined in the vicinity of the nano-particle plasmon resonance. Several disadvantages of this system is that the LDOS is not tunable, and placing a photon emitter at the desired spatial position is  challenging. From a practical viewpoint, one desires a spatial position that is translationally invariant, e.g., near a surface, with an LDOS that can be tuned
in a controllable manner.

In this work we use translationally invariant graphene, which is electronically-tunable, to alter the reservoir for a two-level artificial atom (hereafter referred to as a QD) in a controllable manner. Furthermore, it is known that a graphene support structure consisting of a dielectric layer can play a role in shaping the LDOS \cite{Hanson}, and  so we also consider the effect of a substrate on the resonance fluorescence.
Since the plasmon response of graphene exhibits a strong dependance on bias in the low-mid THz (meV) frequency regime, we model  pump fields and QD excitons at these frequencies \cite{Zibik, Hash}. After obtaining the LDOS properties of the medium, we derive and solve a quantum master equation to demonstrate control over the Mollow triplet \cite{Mollow, Loudon} of a QD by a nearby graphene sheet.
The complex reservoir including the graphene constitutes a lossy inhomogeneous environment for the QD, and here we use a rigorous
photon Green function theory applicable for arbitrary lossy media \cite{Welsch}.
 The Mollow triplet is caused by coherent Rabi oscillations and quantum fluctuations and is of fundamental importance. 
In addition to exploring how the Mollow triplet changes with a tunable graphene layer, we also show that thermal bath effects are important for low THz frequencies.

\begin{figure}[t]
        \centering       
                \includegraphics[width=0.35\textwidth]{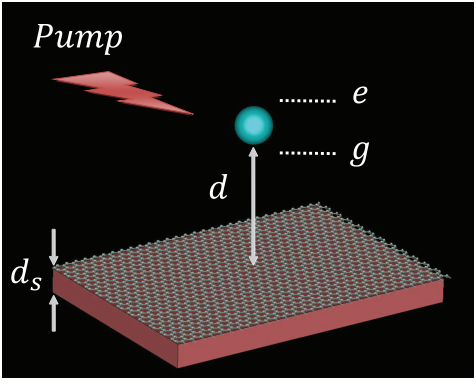}
                \caption{Schematic of a driven QD above supported graphene.} \label{fig:1}
\end{figure}

\section{quantum master equation and Green function theory}

Figure \ref{fig:1} shows the geometry under consideration, consisting of a QD which is a distance $d$ from an infinite graphene sheet. The polarization of the pump field is aligned with the dipole moment, perpendicular to the graphene surface. The Hamiltonian of the coupled system is the sum of QD, pump, reservoir (graphene+vacuum), and their interaction, 
\begin{align}
H_{S}& =\hslash\omega_{x}\sigma^{+}\sigma^{-},
H_{\text{pump}} \!=\!\frac{\hslash\Omega}{2}\left(\sigma^{+}e^{-i\omega_{\text{L}}t}
+\sigma^{-}e^{+i\omega_{\text{L}}t}\right),\nonumber \\
\quad H_{R}& =
\hslash\int d\mathbf{r}\intop_{0}^{\infty}\omega\mathbf{f}^{\dagger}\left(\mathbf{r},\omega\right)
\mathbf{f}\left(\mathbf{r},\omega\right)d\omega, \nonumber \\
\quad H_{I}& =-\left[\sigma^{+}\intop_{0}^{\infty}\mathbf{d}\cdot\mathbf{E}\left(\mathbf{r}_{d},
\omega\right)d\omega+\mathrm{H.c.}\right],
\end{align}
where
 $\omega_{\text{L}}$ is the THz   laser frequency of the pump field, $\sigma^{+}/\sigma_{-}$ are the Pauli operators of the QD exciton, $\Omega=\braket{\mathbf{E}_{\text{pump}}(\mathbf{r}_d)}\cdot \mathbf{d}/\hslash$ is the effective Rabi frequency of the pump source ($\mathbf{d}$ and ${\bf r}_d$ are the QD dipole moment and position),
 $\mathbf{f}^{\dagger}/\mathbf{f}$ are bosonic field operators, $\omega_{x}$ is the exciton resonance, and $\mathbf{E}\left(\mathbf{r}_{d},\omega\right)$ is the total electric field operator at the QD position  \cite{Welsch},
\begin{equation}
\mathbf{E}\left(  \mathbf{r},\omega\right) \! =\!i\sqrt
{\frac{\hslash}{\pi\varepsilon_{0}}}\int\!\sqrt{\operatorname{Im}\!\left(
\varepsilon\left(  \mathbf{r}^{\prime},\omega\right)
\right)  }\mathbf{G}\left(  \mathbf{r},\mathbf{r}^{\prime},\omega
\right)  \cdot\mathbf{f}\left(  \mathbf{r}^{\prime}
,\omega\right)  d\mathbf{r}^{\prime},\label{EF}
\end{equation}
 where the permittivity ($\varepsilon$) and Green function (${\bf G}$) describe the photonic environment (reservoir) of the  graphene and dielectric background. The Green tensor in the quantum field operator  is the classical Green function (propagator) that provides the electromagnetic response at $\mathbf{r}$ due to an excitation at $\mathbf{r}^{\prime}$. All material parameters may be complex-valued. 

This  Hamiltonian is used to form a quantum master equation as described in Ref.~\cite{Vlack}. However, since we are interested in THz operation we do not make the usual zero-temperature bath approximation. Using the traces  
$\mathrm{Tr}_{\text{R}}\![{\mathbf{f}}(\mathbf{r},\omega) 
 {\mathbf{f}}^\dagger (\mathbf{r}^{\prime},\omega^{\prime}) 
 \rho_{R}] =[\overline{n}(\omega)+1]
\delta ( \mathbf{r}-\mathbf{r}^{\prime}) 
\delta ( \omega-\omega ^{\prime}),
$
and
$
\mathrm{Tr}_{\text{R}}\left[{\mathbf{f}}^\dag \left( \mathbf{r}%
,\omega \right) {\mathbf{f}}\left( \mathbf{r}^{\prime },\omega
^{\prime }\right) \rho _{R}\right]=\overline{n}(\omega) 
\delta( \mathbf{r}-\mathbf{r}^{\prime}) 
\delta(\omega -\omega^{\prime})
$, 
where the average thermal photon number is $\overline{n}=\left( e^{\hslash \omega /k_{B}T}-1\right) ^{-1}$ and $\rho _{R}=\rho _{R}\left( 0\right) $ is the density operator of the
reservoir which is assumed to initially be in thermal equilibrium, 
we obtain the master equation for the time-evolution of the density operator
($\rho = \rho(t)$),
\begin{align}
 &\frac{d}{dt}\rho =-\frac{i}{\hslash }\left[ {H}_{\text{S}},\rho
 \right] -\int_{0}^{t}d\tau \Big\{ J_{\rm ph}^{\overline{n}+1}\left( \tau
\right)[ {\sigma }_{+}{\sigma }_{-}\left( -\tau \right) \rho
 \nonumber  \\
 &  - {\sigma }_{-}(-\tau) \rho  {\sigma}_{+}]+\text{H.C.}  \Big\} 
  +L_\text{pure}\,-
 \nonumber
\\
& \!\int_{0}^{t}\!d\tau \left\{ J_{\rm ph}^{\overline{n}}\left( -\tau \right)\left [ {\sigma }%
_{-}{\sigma }_{+}\left( -\tau \right) \rho 
- {\sigma }_{+}\left( -\tau \right)
\rho  {\sigma }_{-}\right ] +\text{H.c.}\right\}  ,
\end{align}

\noindent where $L_{\text{pure}}$ is a pure dephasing term defined in \cite{Vlack}, $\sigma _{\pm }\left( -\tau \right) =e^{-iH_{\text{S}}\tau /\hslash }\widehat{%
\sigma }_{\pm }e^{iH_{\text{S}}\tau /\hslash }$, 
$
\widetilde{J}_{\rm ph}^{\overline{n}}\left( \tau \right) =\int_{0}^{\infty
}d\omega J_{\rm ph}\left( \omega \right) \overline{n}\left( \omega \right)
e^{-i\left( \omega -\omega _{\rm L}\right) \tau },
$
and  the photon reservoir function is related to the Green function through
\begin{equation}
J_\text{ph}\left( \omega \right) =\frac{\mathbf{d}\cdot \text{Im}\left( \mathbf{G}%
\left( \mathbf{r},\mathbf{r},\omega \right) \right) \cdot \mathbf{d}}{\pi
\hslash \varepsilon _{0}}, \label{LDOS}
\end{equation}
which gives a a measure of the QD-environment coupling. Importantly, although at room temperature the average number of phonons at visible
frequencies is negligible, in the THz range $\overline{n}=O(1)$, and so thermal photon effects are required in general.

We assume laterally-infinite graphene modeled as an infinitesimally-thin, local, two-sided surface characterized by a surface conductivity $\sigma$. The Green functions for a graphene sheet at the interface between two dielectrics are given in \cite{Han2008}, and in \cite{Hanson} for graphene on a finite-thickness dielectric support. Considering the graphene sheet in the plane $y=0$, with material described by $\varepsilon _{1}$ for $y>0$ and $\varepsilon _{2}$ for $y<0$, the Green tensor for points in region $n$  is
\begin{equation}
\underline{\mathbf{G}}\left(  \mathbf{r}, \mathbf{r'} \right)=\left(\underline{\mathbf{I}} \, k_{n}^{2}+\mathbf{\nabla\nabla\cdot}\right)  
\left\{  \underline{\mathbf{g}}
^{\text{p}}\left(  \mathbf{r,r}^{\prime}\right)  +\underline{\mathbf{g}}^{\text{s}}\left(  \mathbf{r,r}^{\prime}\right)  \right\},\label{hpt1}
\end{equation}
where $k_{n}=\omega\sqrt{\mu_{0}\varepsilon_{n}}$ is the wavenumber. 

The principle (p) and scattered (s) Green's function components are 
\begin{align}
&\underline{\mathbf{g}}^{\text{p}}\left( \mathbf{r,r}^{\prime }\right)  =%
\underline{\mathbf{I}}\,\frac{e^{\,ik_{1}R}}{4\pi R},
\ \ \underline{\mathbf{g}}^{\text{s}}\left( \mathbf{r,r}^{\prime }\right)  =
\widehat{\mathbf{y}}\widehat{\mathbf{y}}~g_{n}^{\text{s}}\left( \mathbf{r,r}%
^{\prime }\right) + \nonumber
\\
&
\ \ \left( \widehat{\mathbf{y}}\widehat{\mathbf{x}}\frac{%
\partial }{\partial x}
+\widehat{\mathbf{y}}\widehat{\mathbf{z}}\frac{%
\partial }{\partial z}\right) g_{c}^{\text{s}}\left( \mathbf{r,r}^{\prime
}\right) +\left( \widehat{\mathbf{x}}\widehat{\mathbf{x}}~\mathbf{+~}%
\widehat{\mathbf{z}}\widehat{\mathbf{z}}\right) g_{t}^{\text{s}}\left( 
\mathbf{r,r}^{\prime }\right) ,
\end{align}%
where $\underline{\mathbf{I}}$ is the unit dyadic, $k_{\rho}$ is a radial
wavenumber, $p_{n}^{2}=k_{\rho}^{2}-k_{n}^{2}$, $r=\sqrt{\left(  x-x^{\prime}\right)
^{2}+\left(  z-z^{\prime}\right)^{2}}$, and $R=\left\vert \mathbf{r-r}%
^{\prime}\right\vert =\sqrt{\left(  y-y^{\prime}\right)  ^{2}+r^{2}}$. The Sommerfeld integrals are
\begin{equation}
g_{\beta}^{\text{s}}\left(  \mathbf{r,r}^{\prime}\right)  =\frac{1}{2\pi}%
\int_{-\infty}^{\infty}C_{\beta}\frac{H_{0}^{\left(  1\right)  }\left(
k_{\rho} r\right)  e^{-p\left(  y+y^{\prime}\right)  }}{4p}k_{\rho
}dk_{\rho},
\end{equation}
where $\beta=t,n,c$ depends on the graphene and dielectric support layers . For a pump polarized perpendicular to the graphene surface we only need $G_{zz}$ and $\beta=n$, with
\begin{equation}
C_{n}=\frac{\left( \frac{\varepsilon _{2}}{\varepsilon _{1}}%
p_{1}-p_{2}\right) i\omega \varepsilon _{1}-\sigma p_{1}p_{2}}{\left( \frac{%
\varepsilon _{2}}{\varepsilon _{1}}p_{1}+p_{2}\right) i\omega \varepsilon
_{1}-\sigma p_{1}p_{2}} . \label{C}
\end{equation}%
\noindent For more complex geometries, such as graphene on a multi-layered dielectric, only the coefficient $C_{n}$ changes. 

The wave parameter $p_{n}=\sqrt{k_{\rho}^{2}-k_{n}^{2}}$, leads to
branch points at $k_{\rho}=\pm k_{n}$, and thus the $k_{\rho}$-plane is a
four-sheeted Riemann surface. The standard hyperbolic branch cuts
\cite{Ishimaru} that separate the proper sheet (where $\operatorname{Re}%
\left(  p_{n}\right)  >0$, such that the radiation\ condition as $\left\vert
y\right\vert \rightarrow\infty$ is satisfied) and the improper sheet
are the same as in the absence of surface conductivity $\sigma$. The zeros of
the denominators of $C_{\beta}$ lead to pole singularities in the spectral plane associated with surface plasmon polaritons (SPPs).
Using complex-plane analysis, the scattered Green's function can be written as discrete pole (SPP) contributions plus a branch cut integral over the continuum of radiation modes. For $\varepsilon _{1}=\varepsilon _{2}=\varepsilon$, setting the denominator of (\ref{C}) to zero leads to the (TM) SPP wavenumber
$
k_{\rho }=k\sqrt{1-\left( \frac{2}{\sigma \eta }\right) ^{2}},  \label{tmsw}
$
where $\eta =\sqrt{\mu _{0}/\varepsilon }$. In this case, the vertical wavenumber parameter in the Sommerfeld integrals
becomes $p=\sqrt{k_{\rho }^{2}-k^{2}}=i2\omega \varepsilon /\sigma $ and if $\sigma $ is real-valued then $\mathrm{Re}\left( p\right) >0$ is violated and
the TM mode is on the improper Riemann sheet. Assuming complex-valued
conductivity $\sigma =\sigma ^{\prime }+i\sigma ^{\prime \prime }$, 
%
$
p=\frac{i2\omega \varepsilon }{\sigma }=\frac{2\omega \varepsilon }{%
\left\vert \sigma \right\vert ^{2}}\left( \sigma ^{\prime \prime }+i\sigma
^{\prime }\right) ,
$
and therefore if $\sigma ^{\prime \prime }>0$ (as shown below, when the intraband conductivity
dominates) the mode is a surface wave on the proper sheet, whereas if $%
\sigma ^{\prime \prime }<0$ (interband conductivity dominates) the mode is
on the improper sheet, assuming an $\text{exp}(-i \omega t)$ reference \cite{MZPRL2007,Han2008}. Assuming the dipole moment is perpendicular to the graphene surface, only TM SPPs can be excited.

The graphene surface conductivity is \cite{GSC2007},%
\begin{align}
\sigma \left( \omega \right) =& \frac{ie^{2}k_{B}T}{\pi \hslash ^{2}\left(
\omega +i\gamma \right) }\left( \frac{\mu _{c}}{k_{B}T}+2\ln \left( e^{-%
\frac{\mu _{c}}{k_{B}T}}+1\right) \right)    \nonumber \\
&\!\! + \frac{ie^{2}\left( \omega +i\Gamma \right) }{\pi \hslash ^{2}}%
\int_{0}^{\infty }\frac{f_{d}\left( -\varepsilon \right) -f_{d}\left(
\varepsilon \right) }{\left( \omega +i\Gamma \right) ^{2}-4\left(
\varepsilon /\hslash \right) ^{2}}d\varepsilon ,   \label{gen}
\end{align}
where $\mu_{c}$ is chemical potential, $\gamma$ and $\Gamma$ are phenomenological intraband and interband scattering rates, respectively  ($\tau=1/\gamma$ is the scattering time), $e$ is the charge of an electron, and $f_{d}\left(  \varepsilon\right)  =\left(  e^{\left(\varepsilon-\mu_{c}\right)  /k_{B}T}+1\right)  ^{-1}$ is the Fermi-Dirac distribution. The first and second terms in the conductivity are due to intraband and interband contributions, respectively. For $k_{B}T\ll\left\vert \mu_{c}\right\vert,\hslash\omega$ \cite{GSC2007sum}%
\begin{equation}
\sigma \left( \omega \right) = \frac{ie^{2} \left\vert \mu_{c}\right\vert}{\pi \hslash\left(
\omega +i\gamma \right) } + \frac{ie^{2}
}{4\pi\hslash}\ln\left( \frac{2\left\vert \mu_{c}\right\vert -\left(
\omega+i\Gamma\right)  \hslash}{2\left\vert \mu_{c}\right\vert +\left(
\omega+i\Gamma\right)  \hslash}\right)  .   \label{T0}
\end{equation}
\noindent In the following we use (\ref{gen}) for $T=300$ K and (\ref{T0}) for $T=0$ K calculations. We consider a local (momentum independent) conductivity since the main effect considered here is the nontrivial DOS provided by the graphene plasmon energy dispersion.
The Drude form of the conductivity has been verified in the far-infrared \cite{Daw}-\cite{Kim}, and in the near infrared and visible the interband behavior has been verified in \cite{Li}. In the absence of scattering and bias the high-frequency optical conductivity is $\sigma=\sigma_\textrm{min}=e^2/4\hslash$, which has been verified in optical experiments \cite{Nair}. 

Absorption is associated with both scattering and interband transitions. Since realistic values of $\Gamma$ will have a negligible effect on the results we will ignore interband scattering. For $\hslash \omega < 2\left\vert \mu_{c}\right\vert $ interband absorption is blocked, otherwise interband absorption will often dominate Re($\sigma$).
For the intraband term the value of $\gamma$ generally depends on temperature via phonon interactions, the method of growth/fabrication (e.g., epitaxial, chemical vapor deposition, exfoliation), the presence of impurities, and the presence of a substrate. Measured values of scattering times at room temperature  ranged from a few fs \cite{Daw}-\cite{Choi} to several tens of fs \cite{Daw}-\cite{Tan} to several hundred fs ($\sim$ 0.35 ps \cite{Lee}-\cite{Kim}), and at low temperature scattering times on the order of a few ps have been measured (1.1 ps \cite{Li} and 5 ps  \cite{Tan}). Short scattering times are usually associated with impurities and defects since the room-temperature electron-phonon scattering time is estimated to be a few ps \cite{Berger}. In the following we assume $\tau=5$ ps for $T=0$ and $\tau=0.35$ ps for room temperature results. We assume lossless non-dispersive dielectrics to focus on graphene's electrodynamic response rather than on the substrate response.


\section{Purcell factors}

\begin{figure}[t]
     \centering
     \subfloat[]{\includegraphics[width=0.23\textwidth]{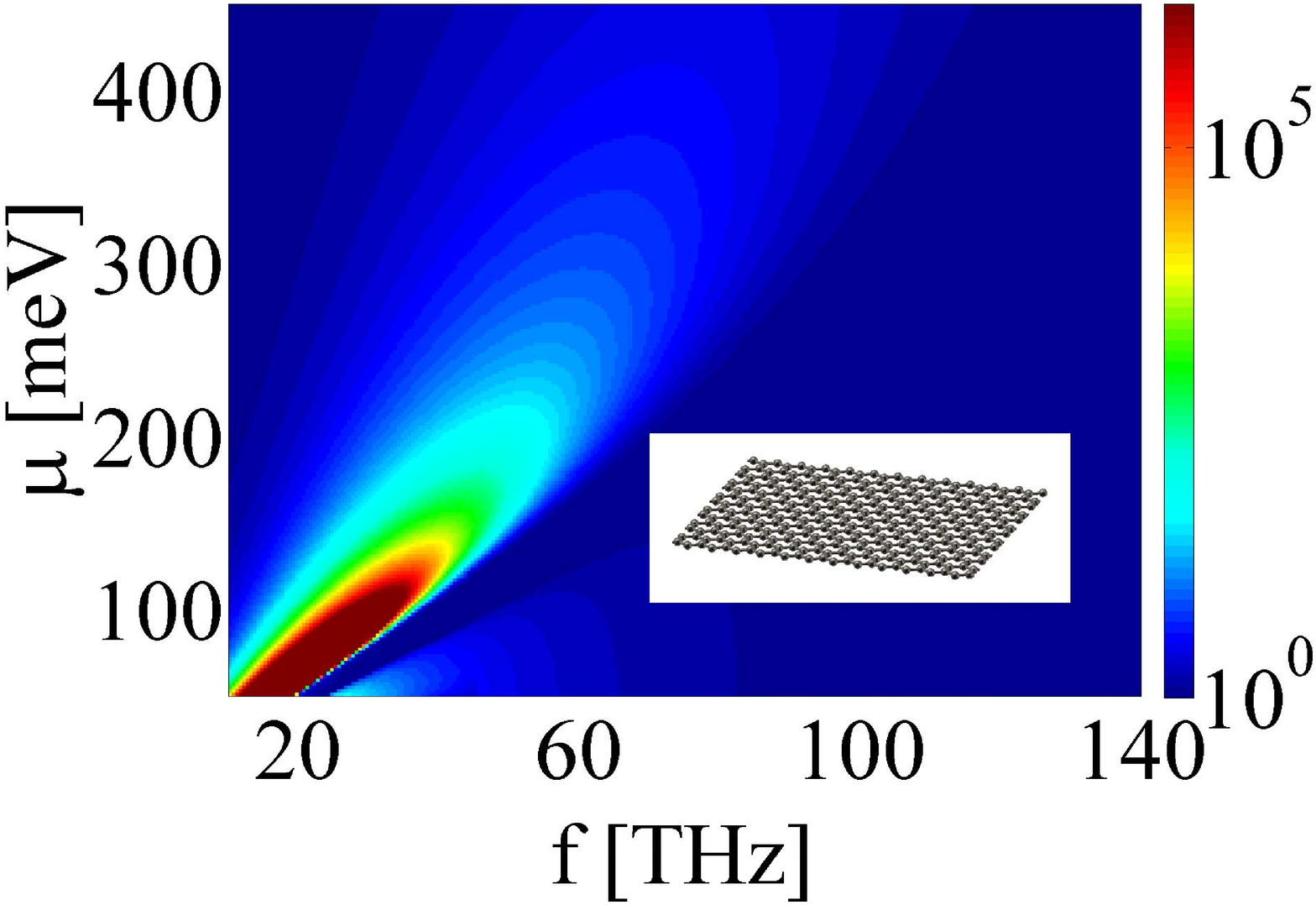}}~
     \subfloat[]{\includegraphics[width=0.23\textwidth]{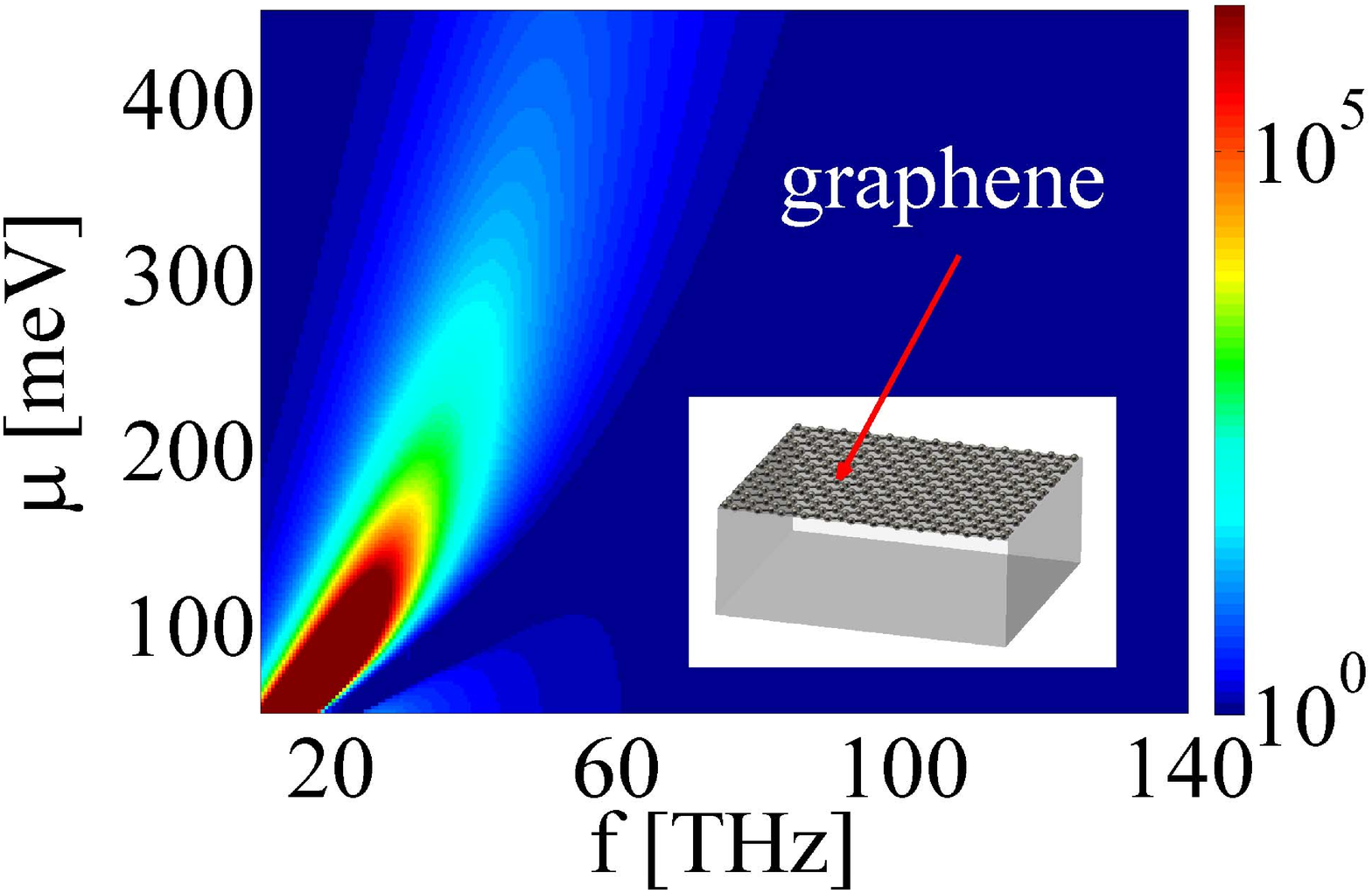}}\vspace{0.02cm}
     
     \subfloat[]{\includegraphics[width=0.23\textwidth]{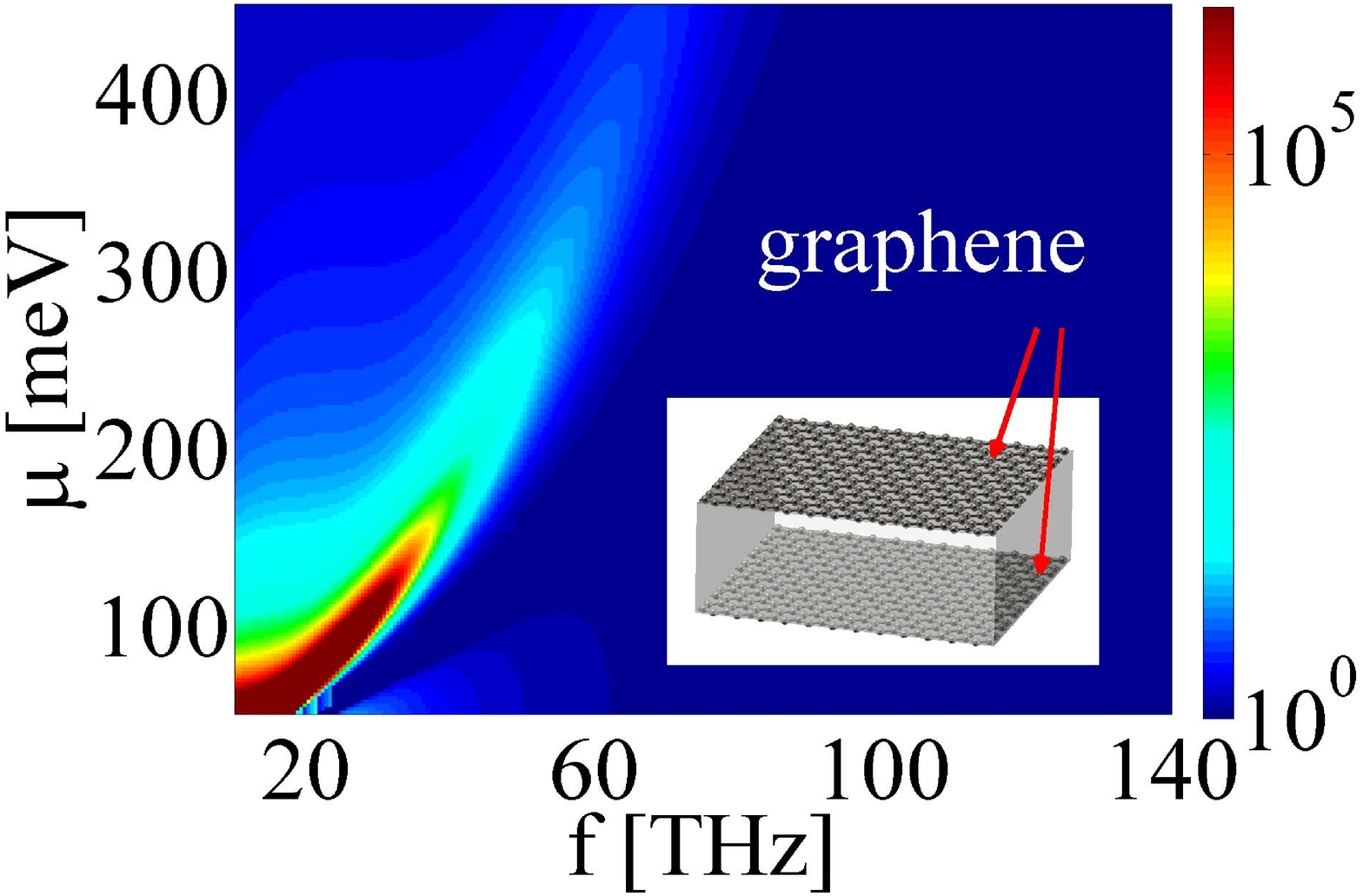}}~
     \subfloat[]{\includegraphics[width=0.25\textwidth]{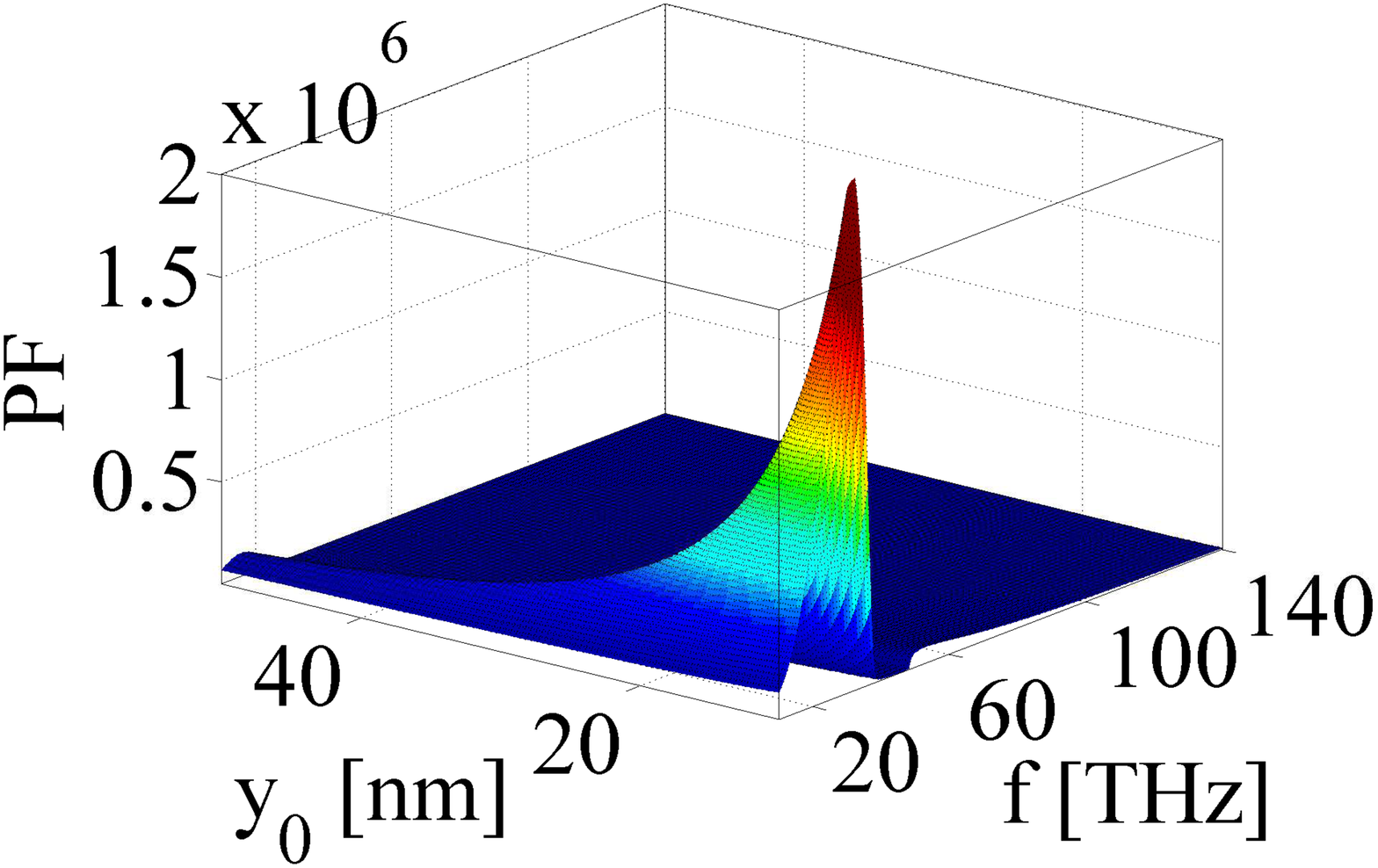}}
     \caption{ (a) The Purcell factor for suspended graphene in vacuum, (b) for graphene on a $d_s=10$ nm, $\varepsilon=4$  substrate, (c) for two layers of graphene separated by a 10 nm, $\varepsilon=4$  substrate, and (d) as a function of position ($y_0$ is the dot position) and frequency for supported graphene on a 10 nm, $\varepsilon=$4  substrate. }
     \label{fig:2}
\end{figure}

The partial LDOS projected normal to the graphene surface, $\rho _{\text{LDOS}}=\left( 6/\pi \omega \right) \mathrm{Im}\left( 
G_{zz}\left( \mathbf{r},\mathbf{r},\omega \right) \right) $, normalized
by the free-space value $\rho _{\text{LDOS}}^{0}=\omega ^{2}/\left( \pi
^{2}c^{3}\right) $ gives the Purcell factor \cite{Purcell} (i.e., the enhanced spontaneous emission factor of a single photon emitter)%
\begin{equation}
\mathrm{PF}=\frac{\rho _{\text{LDOS}}}{\rho _{\text{LDOS}}^{0}}=\frac{6\pi }{k_{0}^{3}}%
\mathrm{Im}\left( G_{zz}\left( \mathbf{r},\mathbf{r},\omega \right)
\right) .
\end{equation}
In the following we consider both suspended graphene, where vacuum exists on either side of the graphene sheet, and supported graphene on a dielectric layer. Figure \ref{fig:2}(a) shows the tunability of the Purcell factor \cite{H} at THz frequencies for a single suspended graphene layer over a range of chemical potentials at a distance of 10 nm from the graphene surface for $\mu_{c}/k_B T\gg 1$ and  $\omega / \gamma \gg 1$ (at room temperature the results of Fig.~\ref{fig:2} will hold with minor quantitative changes). It is clear that in the low THz regime the LDOS and Purcell factor can be tuned considerably by an external bias. Figure \ref{fig:2}(b) shows the Purcell factor for supported graphene on a $d_s=10$ nm thin substrate having relative permittivity $\varepsilon_r=4$ (the approximate permittivity of ${\rm Si0}_{2}$). The presence of the substrate perturbs the SPP of the suspended graphene sheet, and clearly red-shifts the Purcell factor maximums (larger values of $\varepsilon_r$ would further red-shift the Purcell factor). As discussed in \cite{Hanson}, the presence of a substrate tends to confine the SPP mode, leading to higher attenuation as energy concentrates at the lossy graphene surface. 
Figure \ref{fig:2}(c) shows the Purcell factor for two layers of graphene separated by a 10 nm, $\varepsilon_r=4$  substrate. This case closely resembles the result of Fig.~\ref{fig:2}(b), although the bottom graphene layer leads to a parallel-plate like waveguide structure \cite{Hanson_PP}, which tends to further concentrate energy in the dielectric, narrowing and shifting the PF peaks. Figure \ref{fig:2}(d) shows the Purcell factor as a function of position and frequency for supported graphene on a 10 nm, $\varepsilon_r=$4  substrate. Clearly, to be able to tune the QD resonance florescence the QD needs to be  located sufficiently close to the graphene surface to strongly couple to the graphene SPP, due to the strong confinement of the SPP mode. However, one of the advantages of using graphene sheets is that this coupling is translationally invariant in the $x$ and $z$ directions.

\begin{figure}[b]
     \centering
     \subfloat[]{\includegraphics[width=0.23\textwidth]{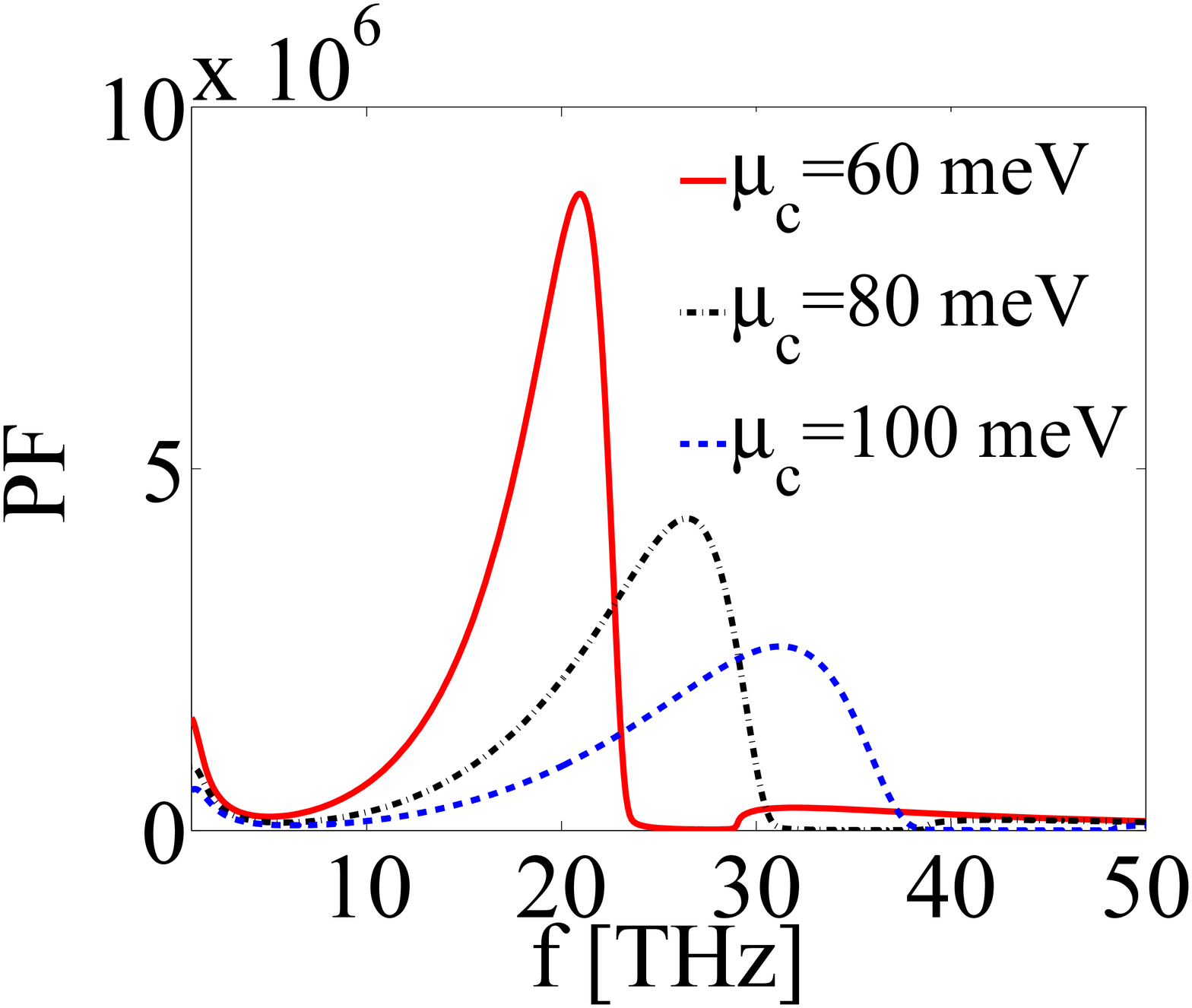}}~
     \subfloat[]{\includegraphics[width=0.23\textwidth]{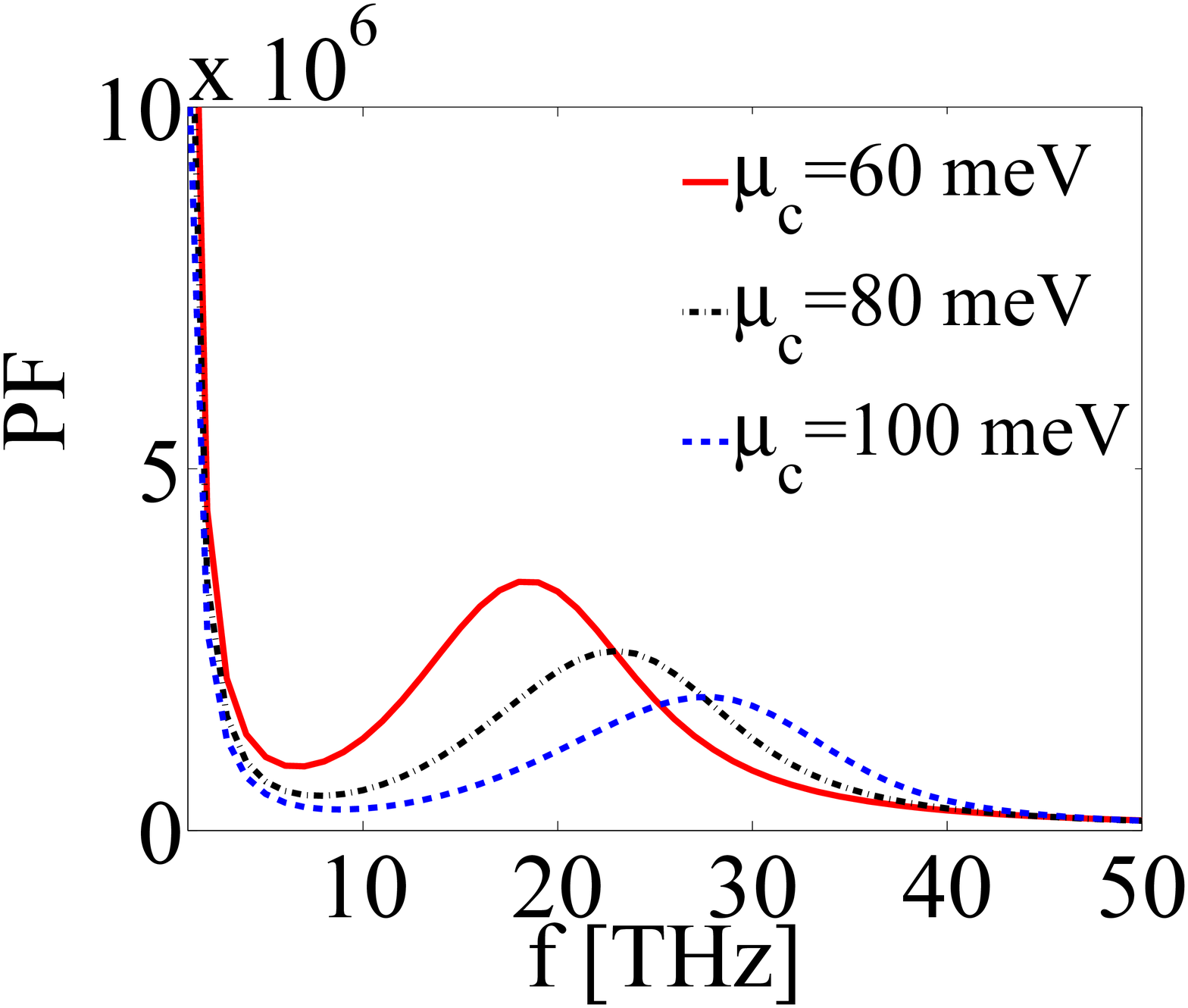}}\vspace{0.1cm}
     \subfloat[]{\includegraphics[width=0.25\textwidth]{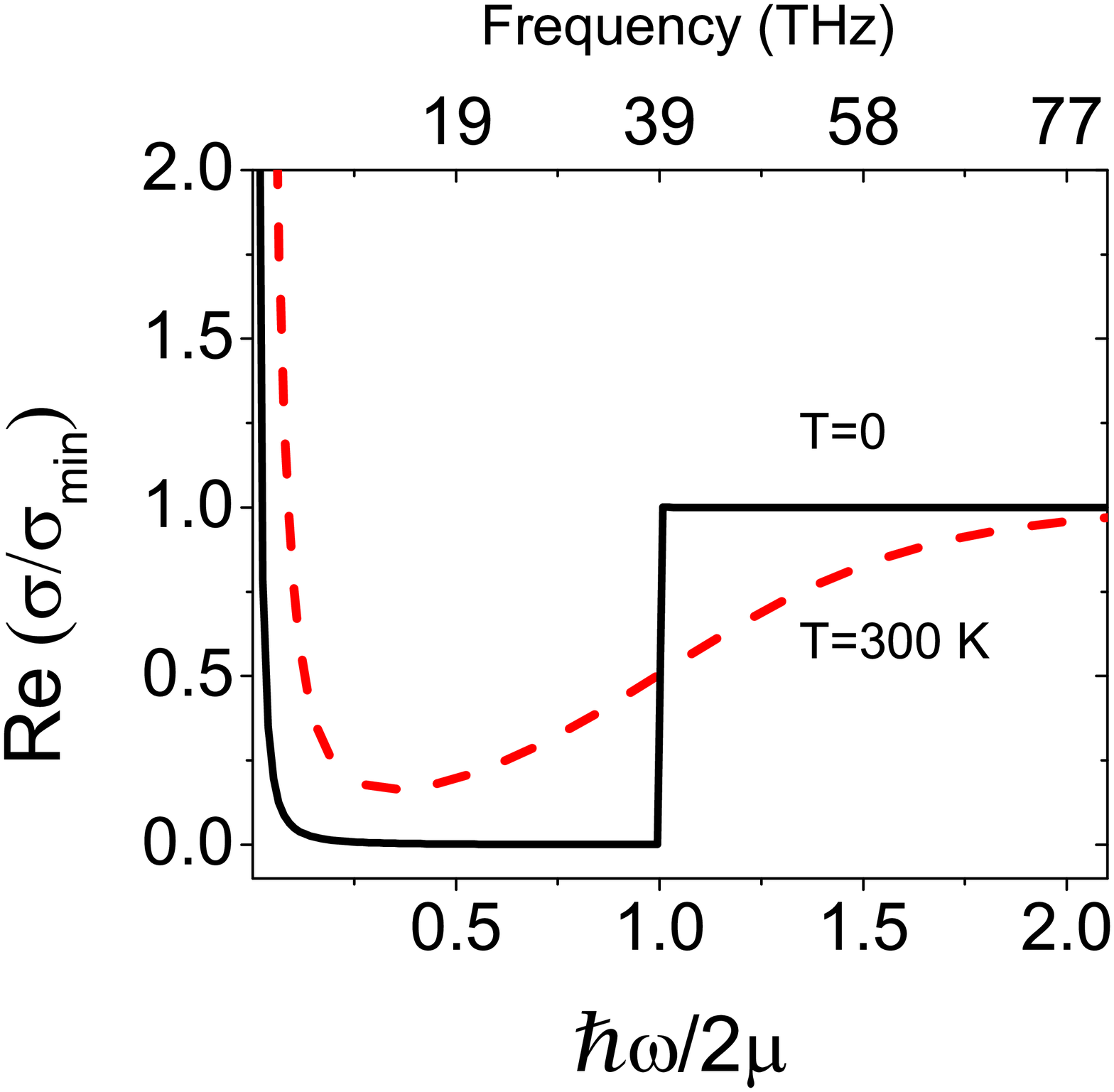}}~
                 \subfloat[]{\includegraphics[width=0.25\textwidth]{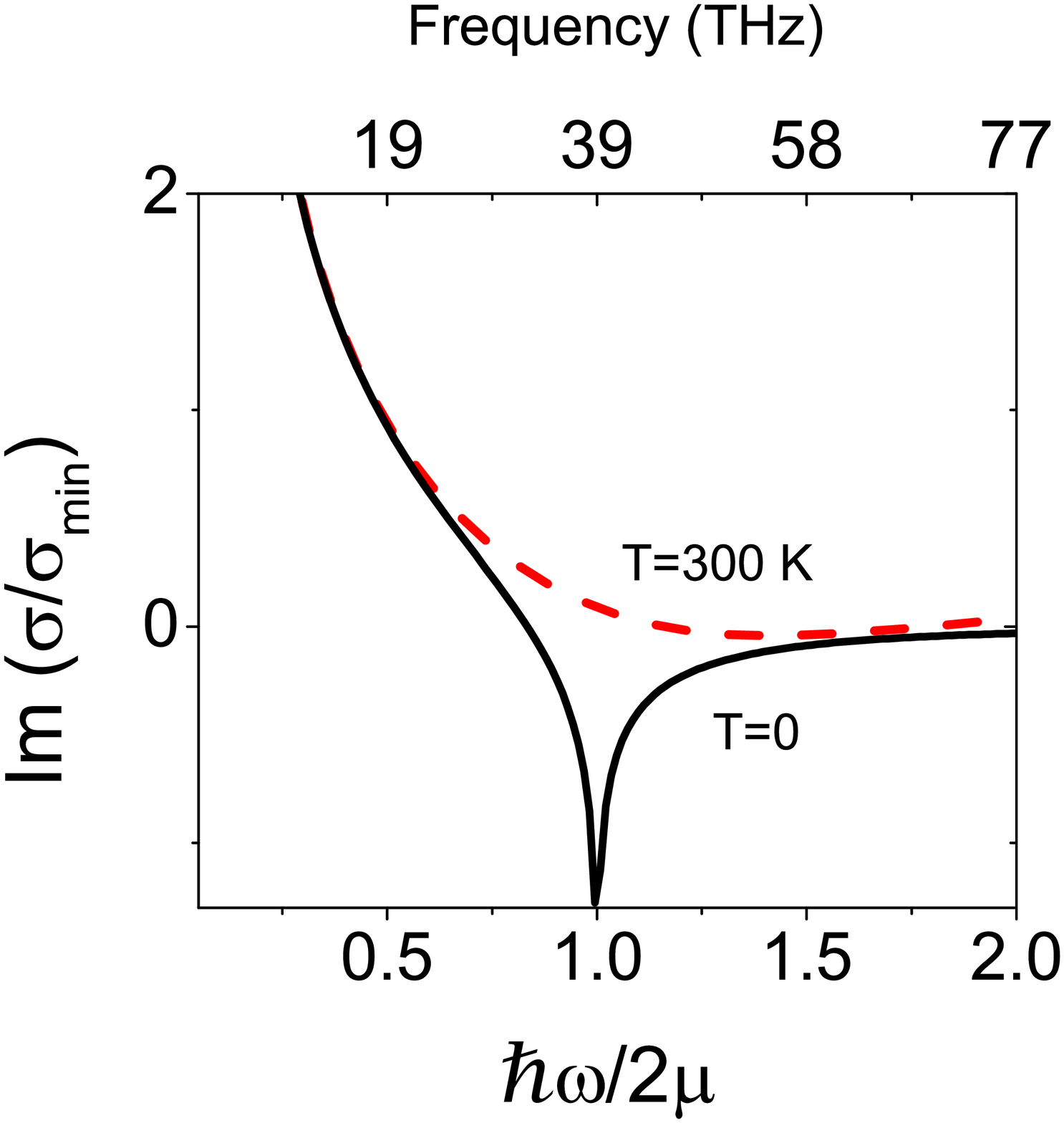}}\vspace{0.02cm}
     \caption{ Purcell factor for suspended graphene for three different chemical potentials at $T=0$ K (a) and $T=300$ K (b). (c)-(d) Real and imaginary part of graphene conductivity at $\mu_c=80$ meV normalized by $\sigma_{\textrm{min}}=e^2/4 \hslash =6.085\times 10^{-5}$ S. }
     \label{fig:3}
\end{figure}

\begin{figure}[b]
     \centering
     \subfloat[]{\includegraphics[width=0.158\textwidth]{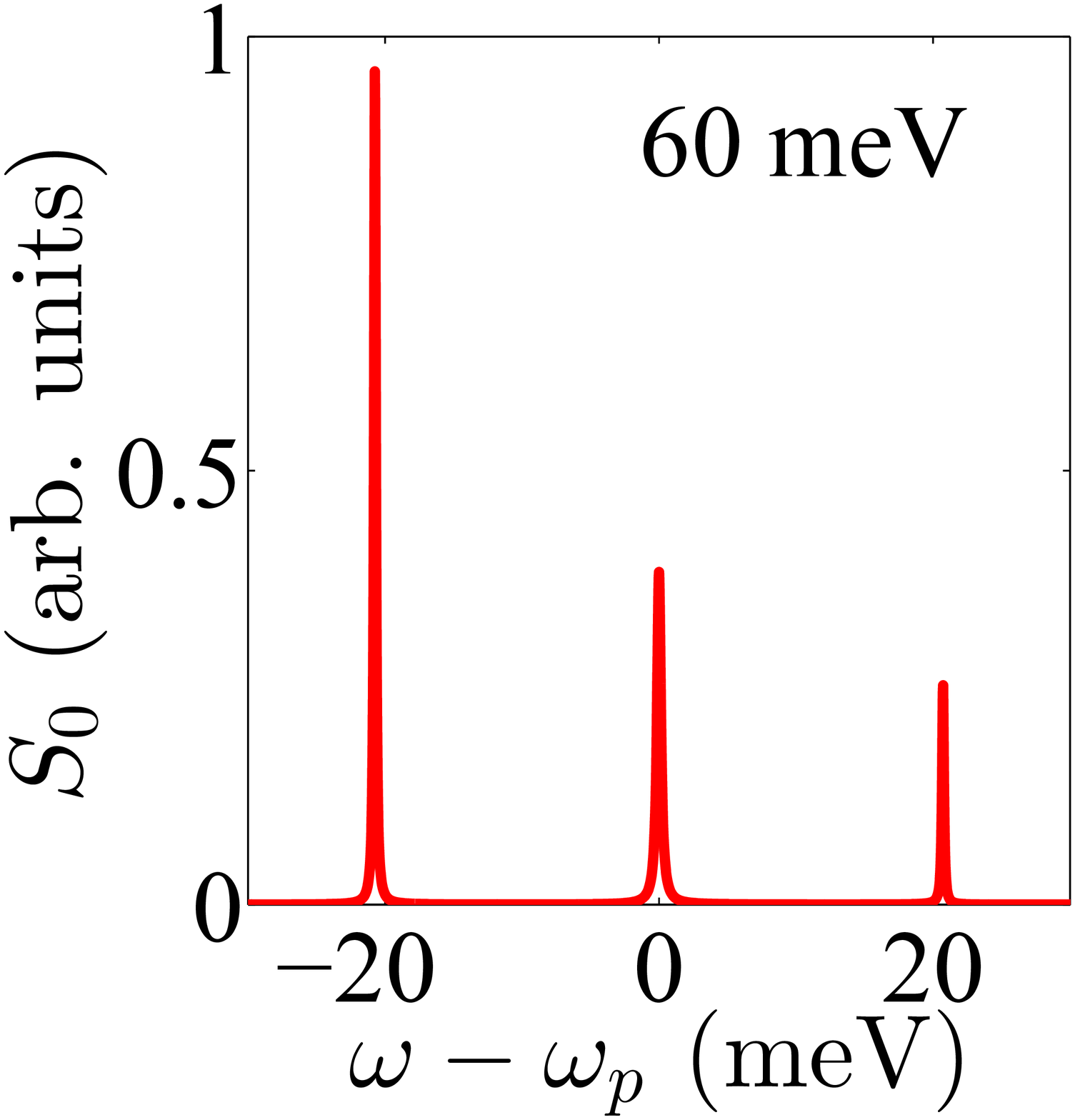}}~
     \subfloat[]{\includegraphics[width=0.158\textwidth]{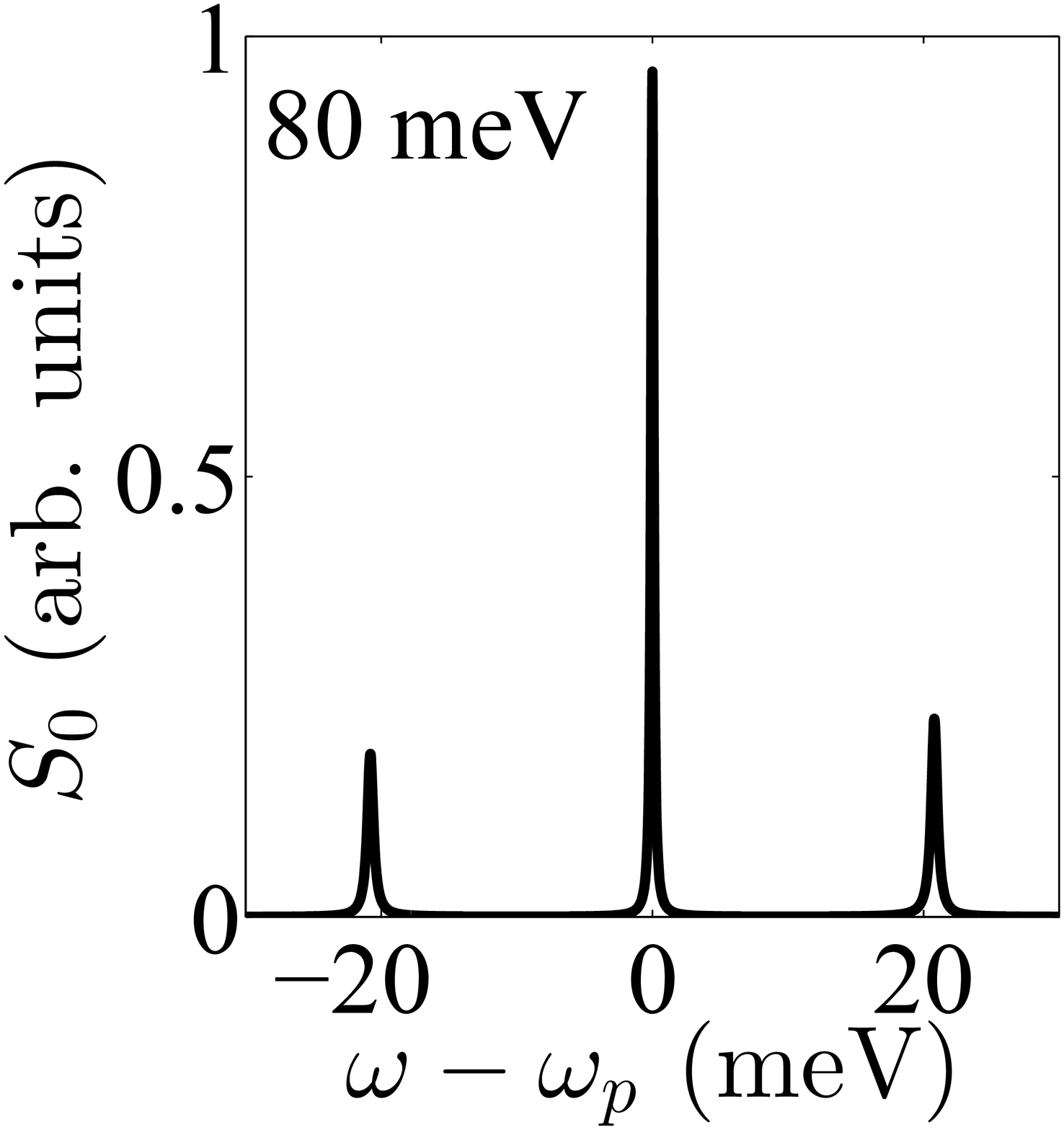}}~
     \subfloat[]{\includegraphics[width=0.16\textwidth]{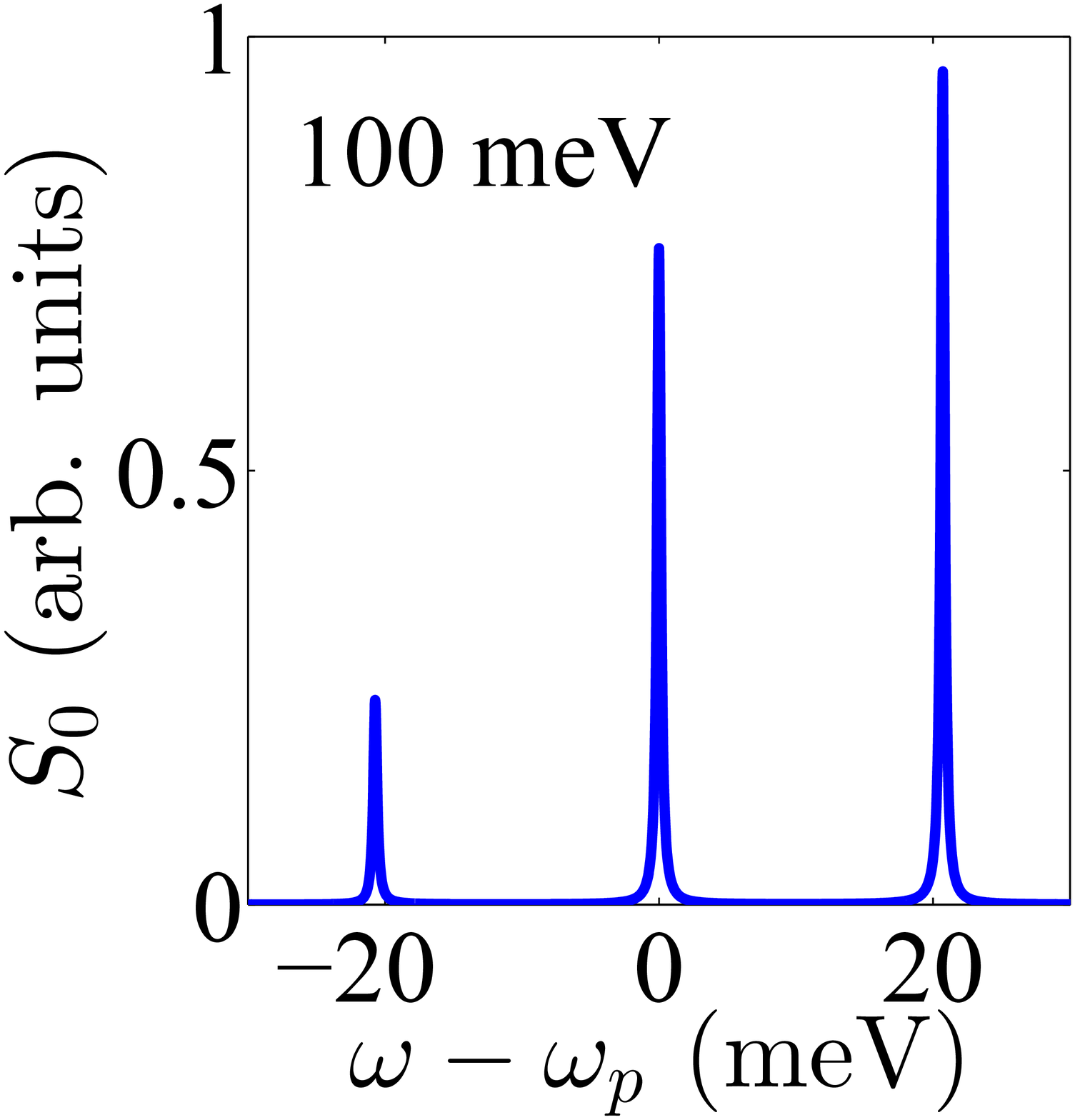}}
     \caption{ Incoherent spectrum of the QD 10 nm above graphene, pumped at 25 THz with (a) $\mu_{c}=60\,$meV,  (b) $\mu_{c}=80\,$meV, (c) $\mu_{c}=100\,$meV.} 
     \label{fig:4}
\end{figure}

Figure \ref{fig:3} shows the Purcell factor for suspended graphene for three different values of chemical potential at $T=0$ (a) and $T=300$ K (b). It can be seen that at the higher temperature the peaks are broadened due to higher absorption compared to the $T=0$ result. Figure \ref{fig:3}(c)-(d) show the conductivity for the $\mu_c=$80 meV case. As frequency increases from a low value, the Drude (intraband) term drops off at $\gamma$ and the interband contribution becomes more important, with a sharp transition in the real-part when $ \alpha=\hslash \omega / 2 \mu_c=$1. The imaginary part undergoes a cusp-discontinuity (for $T=0$ K) at $\alpha=$1. When Im$(\sigma)<0$, as occurs in the vicinity of the cusp, the TM SPP can not propagate (generally, when this occurs a TE SPP can propagate, although for a vertical dipole excitation the TE SPP will not be excited), and this is associated with the drop-off of the Purcell factor. The peak in the Purcell factor corresponds approximately to the frequency where $ \textrm{Im}(\sigma)=\textrm{Re}(\sigma)$.

\section{Spectrum of a driven Quantum Dot}

The incoherent spectrum is defined as 
\begin{align}
S_{0}\left( \omega \right) = \lim_{t\rightarrow 0}\mathrm{Re}&\int_{0}^{\infty
}d\tau  \left\langle \sigma _{+}\left( t+\tau \right) \sigma
_{-}\left( t\right) \right\rangle \nonumber \\
&- \left\langle \sigma _{+}\left( t\right)
\right\rangle \left\langle \sigma _{-}\left( t\right) \right\rangle e^{i\left( \omega _{L}-\omega \right) \tau }d\tau, \label{IS}
\end{align}
    
\noindent where the second term subtracts the coherent scattering from the pump field. The incoherent spectrum of the QD is shown in Figs.~\ref{fig:4}(a-c) for a Rabi frequency of 10 meV at 25 THz, assuming $T=0$ (see the Purcell factor of Fig.~\ref{fig:3}(a)); in all subsequent results the dipole moment is taken to be $30$ Debye. Note that the pump field will naturally be efficiently increased by the coupling to the SPP.  It is evident that by changing the chemical potential of graphene the weights of the sidebands can be substantially changed. That is, the dominant peak of the incoherent spectrum can be shifted, for example, by varying the bias voltage on graphene. As Fig. 3 shows, by varying the bias we can shift the peak of the LDOS; when the LDOS peak aligns with the peak of one of the Mollow triplets, the corresponding triplet is enhanced. Commensurately, a small value of the LDOS at the position of a triplet peak suppresses that peak (due to closure of the plasmon decay channel). This is one of the key results of the paper: the quantum coupling between QDs and graphene can be profoundly influenced by simply changing the bias field. 

Note that in all three cases in Fig.~\ref{fig:4} the exciton-laser detuning is zero ($\omega_{x}=\omega_{L}$) and it is solely the change in the LDOS with bias that is responsible for the significant spectrum tuning. As seen in Fig.~\ref{fig:3}(b), at T=300 K the peaks are broadened but are not significantly shifted compared to the $T=0$ case. Therefore, at room temperature the Mollow triplet can also be controlled. However, in this case since the peaks overlap more than for $T=0$ it is advantageous to choose the Rabi frequency (which controls the separation of the triplet's peaks) and chemical potential values (which control the Purcell factor peaks) to further separate the peaks to achieve similar control over the triplet as in the low temperature case.

The far-field detectable spectrum at position ${\bf r}_D$ is defined as \cite{Vlack}
$S_{p}\left( \mathbf{r}_D,\omega \right) = \frac{2} {\varepsilon_0} \vert \mathbf{d} \cdot \underline{\mathbf{G}}\left(  \mathbf{r}_D, \mathbf{r}_d, \omega \right) \vert ^2 S_0 \left( \omega \right)$.
The factor that multiplies $S_0$ has some features in the vicinity of $\textbf{r}_D=\lambda_\textrm{SPP}$, but is otherwise dominated by the homogeneous-space part of the Green function and is fairly featureless, and so the Mollow triplet of the detectable spectrum will resemble $S_0$.

\begin{figure}[tbh]
     \centering
     \subfloat[]{\includegraphics[width=0.23\textwidth]{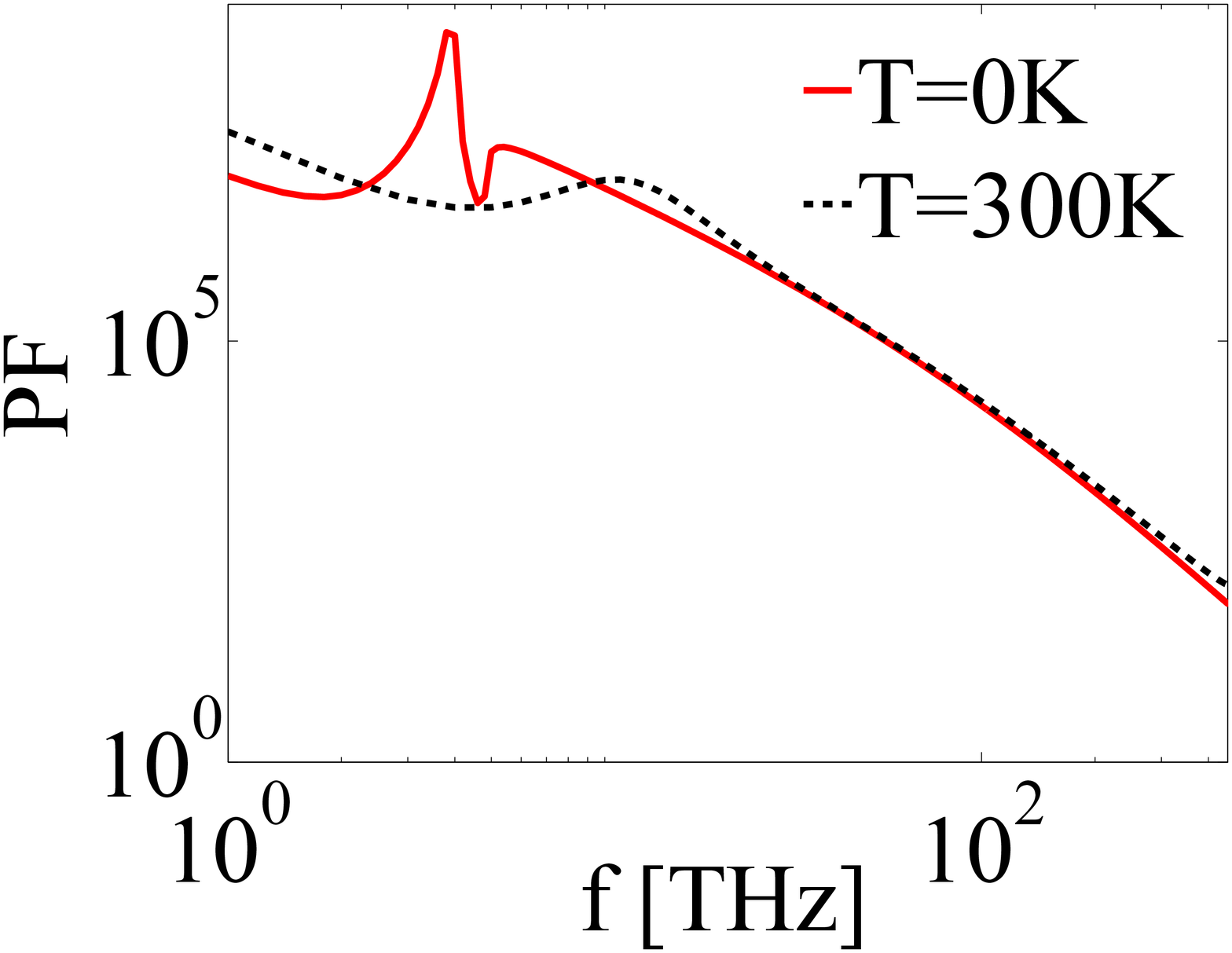}}
     \subfloat[]{\includegraphics[width=0.23\textwidth,height=0.18\textwidth]{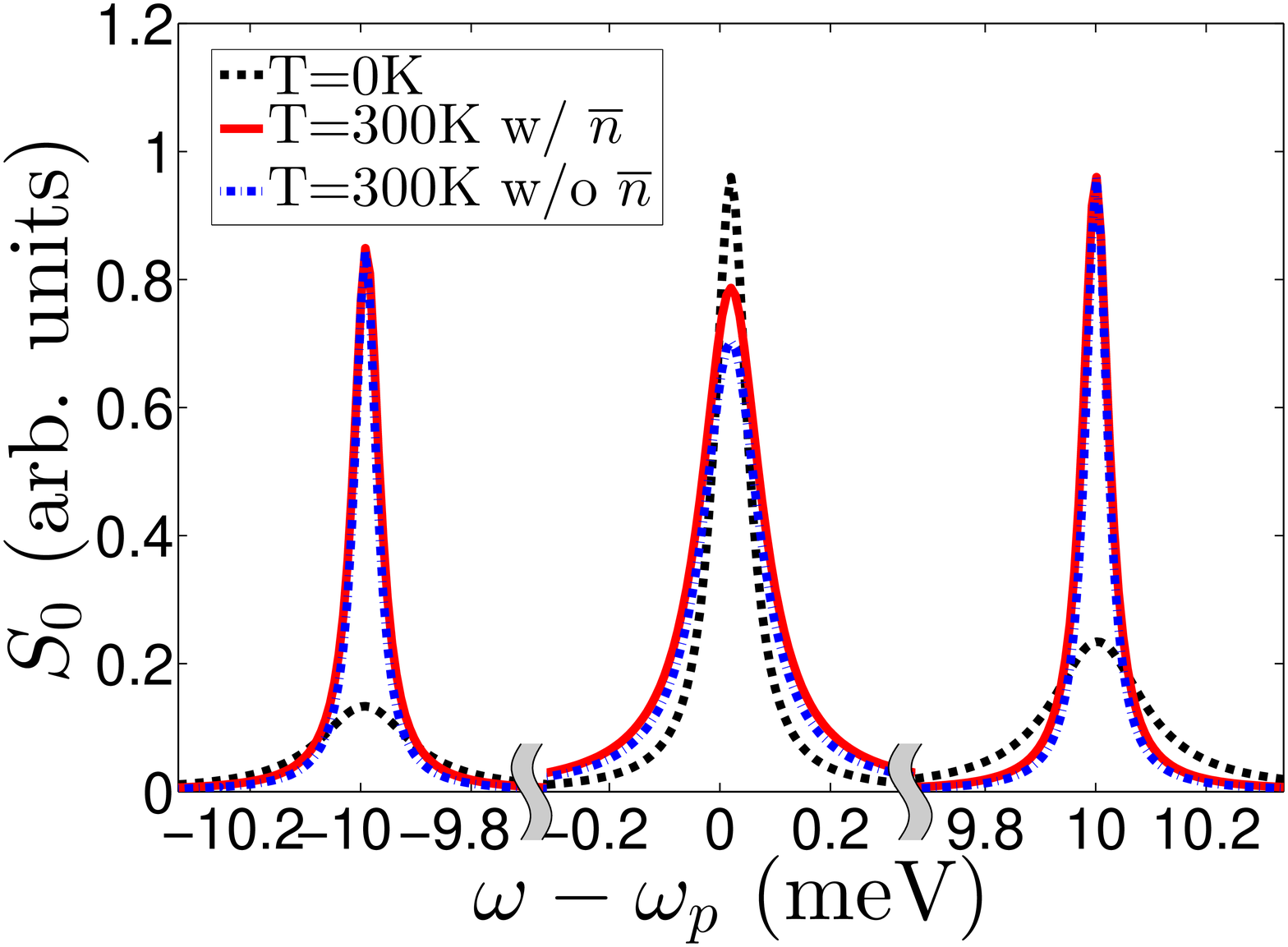}}\vspace{0.1cm}
                \subfloat[]{\includegraphics[width=0.26\textwidth]{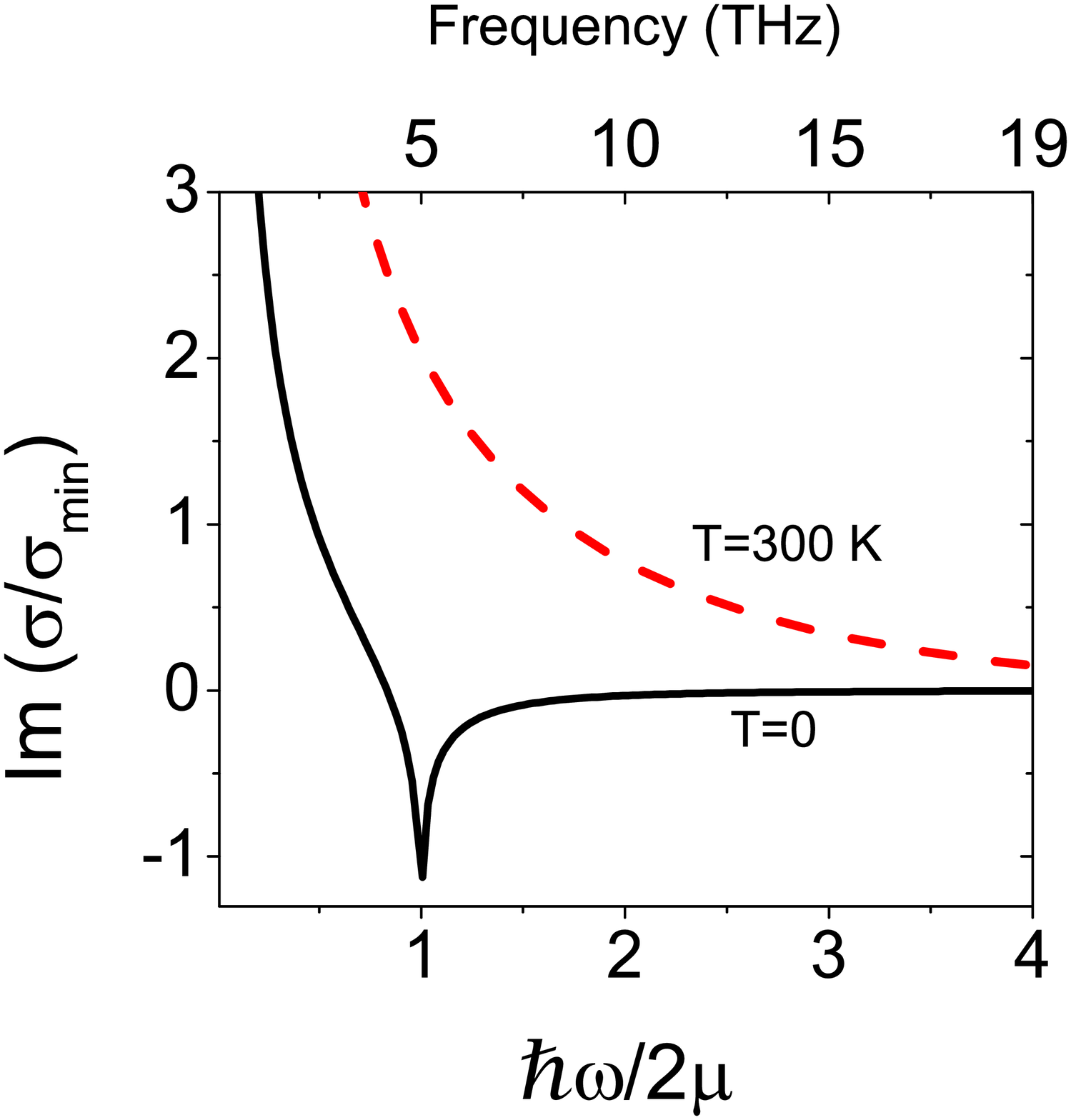}}
     \subfloat[]{\includegraphics[width=0.26\textwidth]{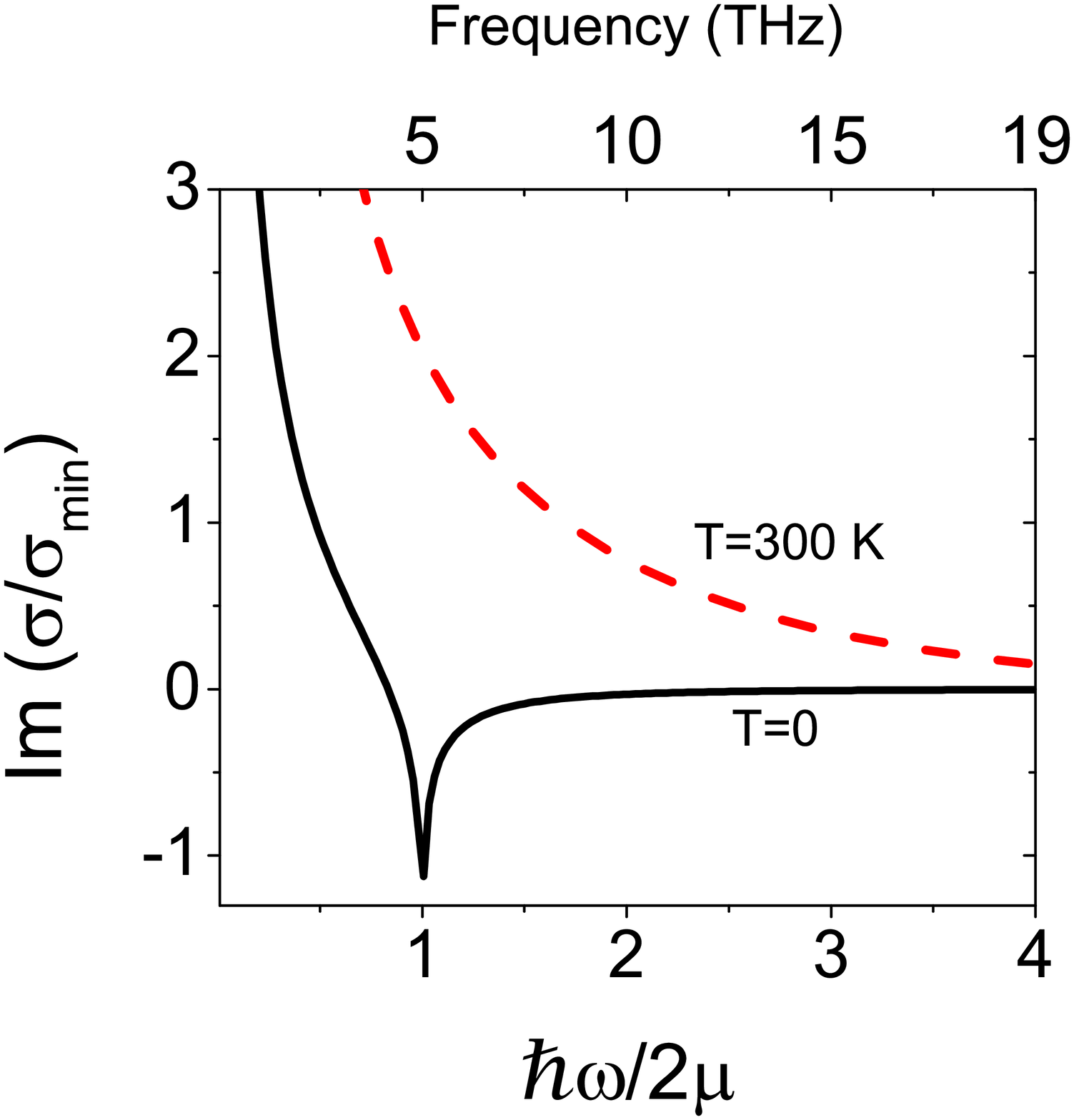}}
     \caption{ (a) Purcell factor for suspended graphene at $x=10\,$nm and $\mu_{c}=10\,$meV at $T=0\,$K and $T=300\,$K, (b) The incoherent spectrum of the QD above graphene for a Rabi frequency of 10 meV and pump resonance at $4$ THz, with and without including $\overline{n}$, (c)-(d) show the corresponding conductivity. }
     \label{fig:5}
\end{figure}

In Fig.~\ref{fig:3} the considered values of $\mu_c$ lead to peaks of the Purcell factor in the range 20-30 THz. For these bias values and frequencies interband absorption is blocked, and the only damping of the conductivity is due to scattering. At several tens of THz, but for room temperature, $\mu_c /k_B T\gg 1$ is at least weakly satisfied, and the Purcell factor results do not change qualitatively, although enhanced absorption broadens the curves (Fig.~\ref{fig:3}(b)). Furthermore, at several tens of THz, even at room temperature the average photon number $\overline{n}$ is negligible, and so the zero-temperature bath approximation holds. However, for small-enough bias the Purcell factor peaks below a few THz, in which case both $\mu_c /k_B T\gg 1$ and $\omega / \gamma \gg 1$ are violated at room temperature. At frequencies of a few THz and for small bias, temperature plays an important role in both the Purcell factor via the graphene conductivity (interband transitions will not be not blocked, leading to enhanced absorption, and the location of PF peak blueshifts due to the second inequality being violated) and in the incoherent spectrum via the effect of the average photon number being non-negligible. To examine this effect, Figure \ref{fig:5}(a) shows the Purcell factor at $x=10\,$nm and $\mu_{c}=10\,$ meV at $T=0\,$K and $T=300\,$K. Figure \ref{fig:5}(b) shows the effect on the resonance fluorescence spectrum. Also shown is the effect of including $\overline{n}$ in the $T=300\,$K calculation, where it can be seen that the inclusion of the room temperature thermal bath is important at these low THz frequencies. Figure \ref{fig:4}(c)-(d) show the normalized conductivity; compared to the 300 K result in Fig.~\ref{fig:3}(c)-(d) for $\mu_c=80$ meV, here the intraband contribution is still important when the interband term becomes active, significantly perturbing the LDOS from the $T=0$ K case (the Drude fall-off is set by $\tau$ and the onset of interband absorption is set by $\mu_c$, and so these two effects can be independently controlled, although $\mu_c$ also governs the amplitude of the intraband contribution). Since we keep the same pump frequency for $T=0$ K and $T=300$ K, in the latter case the LDOS is relatively flat at the pump frequency. 
\section{Conclusions}

In conclusion, 
we have shown that the
Purcell effect and the
 Mollow triplet of a two-level emitter can be tuned by varying the chemical potential of a nearby graphene layer.
We have modeled this effect using an exact Green function theory for
the LDOS and exploited  a quantum master equation to model the quantum dynamics.
We have also demonstrated the important influence of temperature.
This  novel QD-graphene system
allows considerable spectral control at the quantum level via altering an easily-assessable external parameters of the system.

\section*{Acknowledgements}
This work was supported by the Natural Sciences and
Engineering Research Council of Canada. We thank Ronchun Ge for useful discussions.

\end{document}